\documentclass{jfm}
\usepackage{graphicx}
\usepackage{epstopdf, epsfig}
\usepackage{mathrsfs,amsmath}
\usepackage[english]{babel}
\usepackage{subcaption}
\usepackage{textcomp}
\usepackage{caption}
\usepackage{hyperref}

\pdfoutput=1

\usepackage[dvipsnames]{xcolor}

\shortauthor{C. Gadal, C. Narteau, S. Courrech du Pont, O.Rozier and P. Claudin}

\title{Incipient bedforms in a bidirectional wind regime}

\author{
Cyril Gadal\aff{1}\corresp{\email{gadal@ipgp.fr}},
Cl\'ement Narteau\aff{1},
Sylvain Courrech du Pont\aff{2},
Olivier Rozier\aff{1}
\and
Philippe Claudin\aff{3},
}

\affiliation{
\aff{1}
Institut de Physique du Globe de Paris, Sorbonne Paris Cit\'e,
Univ.~Paris-Diderot, UMR 7154 CNRS,
1 rue Jussieu, 75238 Paris Cedex 05, France.
\aff{2}
Laboratoire Mati\`ere et Syst\`emes Complexes,
Sorbonne Paris Cit\'e, Univ.~Paris Diderot,
UMR 7057 CNRS, B\^atiment Condorcet,
10 rue Alice Domon et L\'eonie Duquet, 75205 Paris Cedex 13, France.
\aff{3}
Physique et M\'ecanique des Milieux H\'et\'erog\`enes,
PMMH UMR 7636 CNRS -- ESPCI PSL Research University -- Sorbonne universit\'e --  Sorbonne Paris Cit\'e,
10 rue Vauquelin, 75005 Paris, France.
}

\begin{document}

\maketitle

\begin{abstract}
{\color{blue}
An edited version of this paper was published by Cambridge University Press:
\begin{itemize}
	\item Gadal, C., Narteau, C., Courrech du Pont, S., Rozier, O., \& Claudin, P. (2019). Incipient bedforms in a bidirectional wind regime. Journal of Fluid Mechanics, 862, 490-516. doi:10.1017/jfm.2018.978.
\end{itemize}

To view the published open abstract, go to \href{https://doi.org/10.1017/jfm.2018.978}{https://doi.org/10.1017/jfm.2018.978}.
}
\vspace{.5cm}

Most terrestrial sand seas form at `horse' latitudes, where the wind direction exhibits seasonal variation. Here, we extend the two-dimensional linear stability analysis of a flat sand bed associated with a unidirectional wind to the three-dimensional case in order to account for multidirectional wind regimes. Focusing on the simplest case of bidirectional flow regimes, we show that the transition from transverse to oblique or longitudinal patterns is controlled by the transport ratio and the divergence angle between the two flows. Our predictions agree with previous results for dune orientation, and also provide a wider range of possible alignments depending on flow strength, especially when the two winds are perpendicular, at which the transition occurs. This analysis also predicts the selected pattern wavelength, which either decreases close to the transition angle for strong winds, due to a geometric effect, or increases at low winds, when the bed slope affects the transport. This theoretical analysis is complemented by analogous subaqueous experiments, where bedforms are submitted to alternate water flows. For transverse bedforms, the experimental data validate the model at strong flows, providing evidence for the predicted geometric effect, but also for the increase of the wavelength close to the transport threshold. For longitudinal bedforms, a discrepancy is observed, which we interpret as the sign of enhanced nonlinearities induced by the development of slip faces when the flow alternately blows on both sides of the dune.
\end{abstract}


\section{Introduction}

A turbulent flow over an erodible granular bed can produce various bedforms resulting from the interaction between the flow, the bed topography and the moving sediment. Documented in all arid deserts on Earth, aeolian dune patterns have been historically classified according to their shape and orientation \citep{Fryb79,Wass83,Hunt83,bPye,bLanc}. Under a unidirectional wind, dune crests are perpendicular to the flow and a decrease in sand availability gives rise to a transition from transverse linear dunes to crescentic barchans \citep{Wils73,Main84}. Both have an asymmetric elevation profile with a gentle stoss slope and an avalanche slip face on the lee side. In areas exposed to highly variable wind directions for which the resultant transport is small, star dunes with multiple crest orientations and arms pointing in different directions develop \citep{Lanc89,Zhan12}. However, given the seasonal and inter-annual climate variability on Earth, most sand seas are subjected to a pair of dominant winds and exhibit dune fields composed of long parallel sand ridges with a symmetric shape, a regular spacing and a clear orientation \citep{Lanc82}. Interestingly, this orientation can be transverse, oblique or longitudinal with respect to the resultant sand transport direction \citep{Hunt83,Ping14}.

Despite the wind variability in natural environments, dune morphodynamics in multidirectional wind regimes has been the subject of relatively few studies exploring only a limited number of configurations, partly due to the difficulties of implementing experimental and numerical methods \citep{Rubi87,Rubi90,Wern95,Zhan05,Part09,Reff10,Tani12,Zhan12, Gao18}. Recently, \citet{Cour14} established that two distinct dune orientations exist depending on sand availability. These orientations result from two independent dune growth mechanisms \citep{Luca14,Luca15,Gao15,Lu17}. In zones of high sediment availability, dunes align to maximise the gross sand transport normal to the crest \citep{Rubi87,Rubi90,Cour14,Ping14}.

Over the last 15 years, continuous dune models have been developed and used in unidirectional wind regimes to describe barchan and transverse dunes \citep{Andr02,Kroy02a,Kroy02,Hers04,Hers04a,Kroy05,Schw05,Part07}. Dune dynamics has also been studied at larger scales by cellular automaton models \citep{Nish93,Wern95,Baas06,Nart09,Zhan10,East11,Rozi14,Zhan14} to investigate dune interactions and pattern coarsening \citep{Worm13,Geno13,Gao15a}. In addition to these numerical studies, linear stability analyses of a flat sand bed have also been conducted considering a unidirectional steady flow \citep{Andr02,Elbe05,Koua05,Clau06,Nart09,Four10}. They have shown that the emergence of aeolian dunes (or subaqueous ripples and dunes) results from the balance between a destabilising hydrodynamical process that shifts the maximum basal shear stress upwind of the dune crest \citep{Clau13} and a stabilising transport process that induces a downwind lag of the sand flux. This flux needs a length $L_{\textrm{\emph{sat}}}$ to relax towards its saturated value \citep{Saue01, Andr10,Char13}.

In this work, we extend the linear stability analysis of a flat sand bed to multidirectional wind regimes. This complements the dimensional analysis of \citet{Cour14} by a more complete description of the flow and transport properties. We find that the orientation selected by the most unstable mode agrees with the prediction of \citet{Cour14}. Furthermore, this analysis allows us to predict the corresponding selected pattern wavelength and velocity, which were not addressed previously.

The rest of the paper is constructed as follows. Similarly to the approach of \citet{Deva10} and \citet{Andr12} for subaqueous chevrons and bars, we first consider the interaction between a unidirectional flow and a sinusoidal bed whose crests make an arbitrary angle with respect to the direction of this flow (section~\ref{unimodal}). Combining two alternate winds, we then compute the growth rate of the bed to deduce the orientation, wavelength and velocity of the fastest-growing mode as functions of the wind regime parameters (section~\ref{bimodal}). The comparison of these results with experimental data is done in section~\ref{manip}. Finally, we draw conclusions and perspectives in section~\ref{conclusion}.

\begin{figure}
	\centering
	\includegraphics[width=\linewidth]{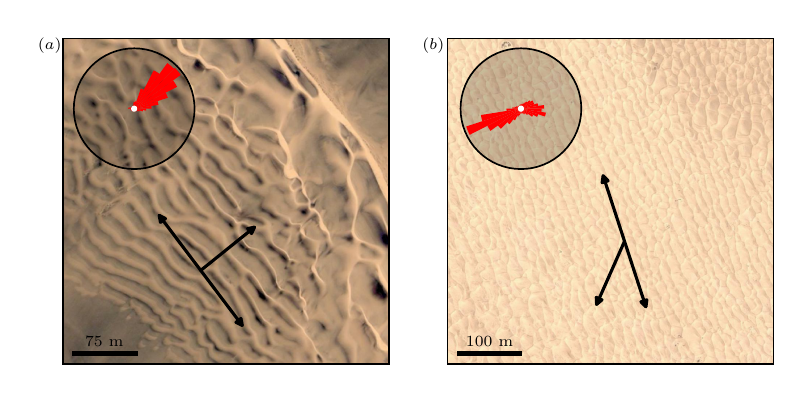}
	\caption{Dune pattern for various multidirectional wind regimes in $(a)$ the Namib desert, Angola ($ 16^\circ 27'$S $11^\circ 58'$E), and $(b)$ the Taklamakan desert, China ($37^\circ25'$N $81^\circ50'$E). Sand flux roses are computed from the wind data of the ERA-Interim project, the latest global atmospheric reanalysis produced by the European Center for Medium-Range Weather Forecasts \citep{Uppa05, Dee11}. Single arrows: corresponding resultant sand flux direction. Double arrows: corresponding prediction of the pattern orientation maximising the growth rate \eqref{Sigma_multi}.}
	\label{Field_figure}
\end{figure}

\section{Linear stability of a flat sand bed in a unidirectional flow}
\label{unimodal}

We consider a unidirectional wind blowing in the $x$-direction above a flat sand bed. The $y$-direction is the transverse coordinate in the horizontal plane and the $z$-direction is vertical. The turbulent air flow is characterised by its fluid density $\rho_f$ and a shear velocity $u_*$, which can be used to determine the bed shear stress $\tau = \rho_f u_*^2$. In the turbulent boundary layer, the wind velocity $u_x(z)$ follows a logarithmic profile,
\begin{equation}
	u_x = \frac{u_*}{\kappa} \ln\left(\frac{z}{z_0}\right),
\label{uxlog}
\end{equation}
where $\kappa = 0.4$ is the von K\'arm\'an constant, and $z_0$ is the aerodynamic roughness of the bed \citep{Schlichting55book}. The aerodynamic roughness of the saltation layer is an increasing function of wind strength \citep{Owen64,Sher92}, and in typical transport conditions is a few times the grain size $d$ \citep{Dura11}. In the presence of aeolian ripples, the corresponding roughness is a fraction of the amplitude of these smaller-scale bedforms. In practice, $z_0$ is of the order of a millimetre. However, it can also reach several millimetres when saltation occurs at high rate.

\subsection{Turbulent flow over a sinusoidal bottom}
\label{flow}

\begin{figure}
	\centering
	\includegraphics[width=\linewidth]{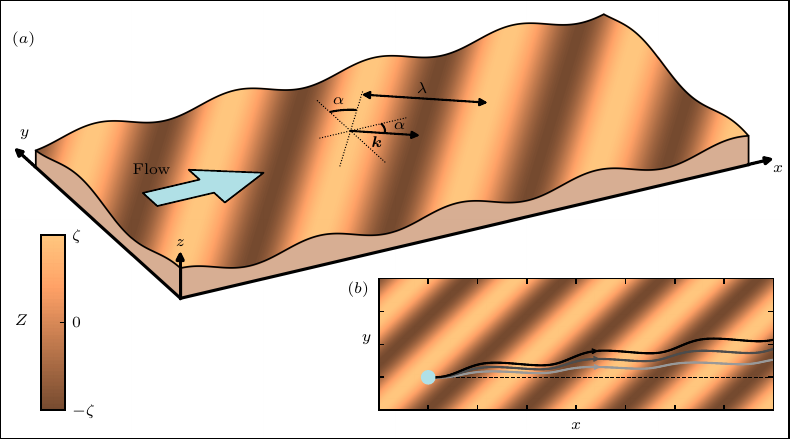}
	\caption{Sinusoidal sand bed. $(a)$ Sketch in a perspective view. The base flow is in the direction of the $x$-axis. The crest lines make an angle $\alpha$ with respect to the transverse $y$-axis. The colour bar indicates the bed elevation. $(b)$ Top view. Solid lines are lines tangent to the transport direction $\boldsymbol{t}$ (here for $k\zeta = 0.25$). Lines from light to dark greys encode increasing values of the velocity ratio $u_*/u_{\textrm{\emph{th}}} \in \{1.6,\, 2.5,\, 5\}$. For reference, the dashed line shows the direction of transport over transverse bedforms ($\alpha = 0^\circ$).}
	\label{schema}
\end{figure}

To investigate the linear stability analysis of the flat bed, we consider a sinusoidal bed perturbation whose elevation profile is of the form
\begin{equation}
	Z = \zeta e^{ik (\cos\alpha x + \sin\alpha y-ct) + \sigma t},
\label{defZ}
\end{equation}
where $\alpha$ is the angle between the wavevector $\boldsymbol{k} = (k\cos\alpha, k\sin\alpha)$ and the flow direction, as sketched in figure~\ref{schema}$(a)$. The corresponding wavelength is $\lambda=2\pi/k$. The amplitude $\zeta$ of the perturbation is such that $k\zeta \ll 1$. Here $\sigma$ is the growth rate and $c$ the propagation velocity of the perturbation. We compute the dispersion relations $\sigma(\alpha,k)$ and $c(\alpha,k)$ following \citet{Andr12}. Denoting by a hat $\hat{.}$ the Fourier transform of the variables, the $x$- and $y$- components of the bed shear stress perturbation induced by the undulated bed \eqref{defZ} can be written as:
\begin{equation}
	\hat{\tau}_{x} = \tau \left(\mathcal{A}_{x}+i\mathcal{B}_{x}\right)k\hat{Z}
	\qquad \textrm{and} \qquad
	\hat{\tau}_{y} = \tau \left(\mathcal{A}_{y}+i\mathcal{B}_{y}\right)k\hat{Z},
\label{defAxBxAyBy}
\end{equation}
where $\{\mathcal{A}_{x},\, \mathcal{A}_{y}\}$ and $\{\mathcal{B}_{x},\, \mathcal{B}_{y}\}$ are the in-phase and in-quadrature components of the stress perturbation with respect to the topography, respectively. These coefficients are functions of the wavenumber, and can be numerically derived from the linearisation of a turbulent model based on the Navier-Stokes equations on a steady sinusoidal bed (see the supplementary material of \citet{Andr12}). However, this dependence can be neglected for unbounded flows, and it is the case for aeolian dunes which are typically much smaller than the atmospheric boundary layer. For the sake of simplicity, we then use the approximate but analytical expressions:
\begin{eqnarray}
	\mathcal{A}_{x} & = & \mathcal{A}_{0} \cos^{2} \alpha,  \label{Axapprox}\\
	\mathcal{B}_{x} & = & \mathcal{B}_{0} \cos^{2} \alpha,  \label{Bxapprox}\\
	\mathcal{A}_{y} & = & \frac{1}{2} \mathcal{A}_{0} \cos\alpha \sin\alpha,  \label{Ayapprox}\\
	\mathcal{B}_{y} & = & \frac{1}{2} \mathcal{B}_{0} \cos\alpha \sin\alpha,  \label{Byapprox}
\end{eqnarray}
where $\{\mathcal{A}_{0},\, \mathcal{B}_{0}\}$ are the coefficients when the dune crests are perpendicular to the flow ($\alpha = 0^\circ$). The dependence of these coefficients on $\alpha$ corresponds to the scaling of the velocity for a potential flow over a sinusoidal topography, i.e. in the outer layer where the fluid inertia is balanced by the pressure gradient \citep{Andr12}. The factor $1/2$ in the transverse direction is due to the variation of the shear stress as the square of the velocity, which is a common approximation for turbulent flows. Here we use $\mathcal{A}_{0} = 3.5$ and $\mathcal{B}_{0} = 2$, as measured in the field on a flat transverse dune \citep{Clau13}.

\subsection{Sediment transport}
\label{sed_trans}

In our model, sediment transport can be characterised by three quantities. The first one is the steady sediment flux on a flat sand bed, also called the saturated flux $q_{sat}$. It is an increasing function of the basal shear stress $\tau$ above a threshold $\tau_{\textrm{\emph{th}}}$. As argued by \citet{Unga87}, the transfer of momentum from the fluid to the saltating grains in the transport layer slows down the flow to the threshold velocity. This results in a linear transport law of the form \citep{Andr04bis,Crey09,Dura11,Ho11}
\begin{equation}
	\boldsymbol{q}_{\textrm{\emph{sat}}} = \Omega \left(\tau - \tau_{\textrm{\emph{th}}}\right) \boldsymbol{t}.
\label{transport_law_tau}
\end{equation}
Here $\Omega$ is a dimensional constant, and $\boldsymbol{t}$ is the saturated flux direction (corresponding to the direction of the flow $\boldsymbol{e}_x$ for a flat sand bed). In subaqueous environments where bedload is the dominant mode of transport, a power law of exponent $3/2$ is reported for the variation of the sediment flux with the bed shear stress \citep{Bagn56, Meye48, Dura12}. However, the results of the linear stability analysis do not qualitatively depend on the exponent of the transport law.

The second quantity is the transport threshold, which can be expressed in terms of a shear stress $\tau_{\textrm{\emph{th}}}$ or a velocity $u_{\textrm{\emph{th}}}$ defined as $\tau_{\textrm{\emph{th}}} \equiv \rho_f u_{\textrm{\emph{th}}}^2$. For a flat bed with a longitudinal slope $\partial_x Z$, this threshold is modified at the linear order in $k\zeta$ as $\tau_{\textrm{\emph{th}}} (1 + \partial_x Z / \mu)$, where $\mu$ is an effective friction coefficient corresponding to the avalanche slope $\mu \simeq \tan(35^\circ) = 0.7$. This slope effect has been calibrated for both saltation \citep{Hard88,Iver99} and bedload transport \citep{Dey03, Fern76, Lois05}.

At the linear order, the transverse slope $\partial_y Z$ has no impact on the threshold value but affects the transport direction. Following \citet{Andr12book}, in analogy with a sliding frictional disc on an inclined surface, particles are driven by the flow but also attracted downwards in the direction of the steepest slope by gravity. This results in a transport direction that reads
\begin{equation}
        \boldsymbol{t} = \boldsymbol{e}_x  + \left(\frac{\tau_y}{\tau} - \frac{1}{\mu_y}\frac{u_{\textrm{\emph{th}}}}{u_{*}} \partial_y Z \right) \boldsymbol{e}_{y}.
\label{slope_effect_direction}
\end{equation}
This behaviour has been tested for bedload transport by \citet{Seki92}, showing that the coefficient $\mu_y$ is also equal to the avalanche slope $\mu$. However, saltating grains may have a different dynamics because they are not in permanent contact with the bed. In contrast with the case of longitudinal slope, there are no available experimental data to calibrate \eqref{slope_effect_direction}. This expression accounts, however, for the correct symmetry and trend with respect to $\partial_y Z$. For the sake of simplicity, we then also take $\mu_y=\mu=\tan(35^\circ)$, but keep in mind that this may overestimate the impact of the bed transverse slope on the direction of the saltation flux.

The flux deviates from the direction of the base flow $\boldsymbol{e}_x$, due to the $y$-component of the shear stress induced by the oblique perturbation. At large flow velocity, this deflection is maximum, but it is reduced close to the transport threshold by the slope effect. We thus show lines following the transport direction for different velocity ratios $u_*/u_{\textrm{\emph{th}}}$ in figure~\ref{schema}$(b)$. At wind velocities far from the threshold, this slope effect is negligible and the lines follow the shear stress direction. The deviation shortens the upwind path and lengthens the downwind path. Then, the time spent in the downwind side is longer, and the associated deviation prevails, leading to larger deviations on the lee side of the ridges. Close to the transport threshold, the slope affects the transport direction, reducing the drift in the lee zones. The corresponding lines are therefore closer to the base line.

The third and last important feature of sediment transport is related to non-homogeneous or non-steady conditions. The actual flux $\boldsymbol{q}$ at a given location is not determined by the local shear stress \eqref{defAxBxAyBy} only because, as soon as there is a change in flow strength or surface property, it  needs some space and time to relax towards its new steady value \citep{Saue01,Andr02a,Dura11,Paht14}. The response time of the topography is much longer than that of the sediment transport so that temporal equilibrium is assumed between $\boldsymbol{q}$ and $\boldsymbol{q}_{\textrm{\emph{sat}}}$. Close to saturation, a first-order linear relaxation process describes the relaxation process, involving a single length scale $L_{\textrm{\emph{sat}}}$. In a vectorial form to account for both $x$- and $y$- components of the flux, it reads
\begin{equation}
	L_{\textrm{\emph{sat}}} \left(\boldsymbol{t} \boldsymbol{\cdot} \nabla \right) \boldsymbol{q} = \boldsymbol{q}_{  sat} - \boldsymbol{q}.
\label{saturated_relax}
\end{equation}
For aeolian saltation, a combination of wind tunnel experiments and field data show that $L_{  sat}$ is of the order of a metre and proportional to the grain/fluid density ratio $\rho_p/\rho_f$, and to the grain size $d$ \citep{Andr10}.

\subsection{Growth rate and most unstable mode}
\label{DispRel}

The evolution of the bed elevation profile is governed by the sediment mass conservation equation (also known as the Exner equation in this context):
\begin{equation}
	\partial_t Z = -\nabla \boldsymbol{\cdot} \boldsymbol{q}.
\label{Exner}
\end{equation}
Here, the sediment flux $\boldsymbol{q}$ is defined as a volumetric flux, i.e. a volume of grains at the bed packing fraction per unit time and unit length, so that the factor in front of the right-hand side of \eqref{Exner} is unity. To deal with dimensionless quantities, we rescale lengths by $L_{  sat}$ and times by $L^2_{  sat}/Q$, where $Q$ is the reference flux $Q = \rho_f\Omega  u_*^2$.

Following \citet{Andr12}, the dispersion relations can be expressed in a dimensionless form as
\begin{eqnarray}
\sigma & = & \frac{k^{2}}{1+(k\cos\alpha)^{2}}\left[b_{x}\cos\alpha + b_{y}\sin\alpha  - k\cos \alpha\left(a_{x}\cos\alpha +a_{y}\sin \alpha \right)\right], \label{growth_rate} \\
c & = & \frac{k}{1+(k\cos\alpha)^{2}}\left[a_{x}\cos\alpha +a_{y}\sin \alpha  + k\cos \alpha\left(b_{x}\cos\alpha +b_{y}\sin\alpha \right)\right], \label{celerity}
\end{eqnarray}
where the coefficients are
\begin{eqnarray}
	a_{x} & = & \mathcal{A}_{x}, \label{coeffax} \\
	b_{x} & = & \mathcal{B}_{x}  - \cos \alpha \frac{1}{\mu}\frac{u_{\textrm{\emph{th}}}^{2}}{u_{*}^{2}}, \label{coeffbx} \\
	a_{y} & = & \left(1-\frac{u_{\textrm{\emph{th}}}^{2}}{u_{*}^{2}}\right)\mathcal{A}_{y}, \label{coeffay} \\
	b_{y} & = & \left(1-\frac{u_{\textrm{\emph{th}}}^{2}}{u_{*}^{2}}\right)\left(\mathcal{B}_{y} - \sin \alpha \frac{1}{\mu}\frac{u_{\textrm{\emph{th}}}}{u_{*}}\right). \label{coeffby}
\end{eqnarray}
%
%
\begin{figure}
	\centering
	\includegraphics[width=\textwidth]{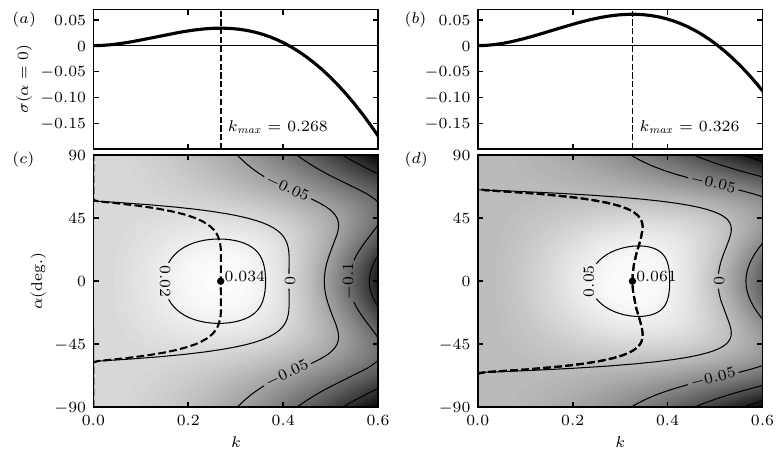}
	\caption{Growth rate $\sigma$ as a function of the wavenumber $k$ and the orientation angle $\alpha$ of the bed perturbation, for $u_*/u_{\textrm{\emph{th}}} = 1.6$ ($a,c$) and $u_*/u_{\textrm{\emph{th}}} = 2.5$ ($b,d$). ($a,b$) Show the reference transverse case $\alpha = 0 ^\circ$. ($c,d$) Show the growth rate in the plane $(\alpha,k)$ with contour lines as indicated. Dashed line: location of the maximum growth rate for each $\alpha$. Black dot: global maximum growth rate, corresponding to the most unstable mode $(\alpha_{\textrm{\emph{sat}}},k_{\textrm{\emph{sat}}})$. Here, for a unidirectional wind regime, $\alpha_{\textrm{\emph{sat}}} = 0^\circ$, independently of $u_*/u_{\textrm{\emph{th}}}$.}
	\label{Sigma_unidirectional}
\end{figure}
%
Plugging the stress coefficients (\eqref{Axapprox}-\eqref{Byapprox}) into the expression for the growth rate, we obtain
\begin{eqnarray}
\sigma & = & \frac{k^{2}}{1+(\cos\alpha k)^{2}} \Bigg[ \cos\alpha\left(\mathcal{B}_{0}- k\cos\alpha\mathcal{A}_{0}\right)\left(\cos^{2} \alpha+ \frac{r}{2}\sin^{2} \alpha\right)
\nonumber \\
& - & \frac{1}{\mu}\frac{u_{\textrm{\emph{th}}}}{u_{*}}\left(\frac{u_{\textrm{\emph{th}}}}{u_{*}}\cos^{2} \alpha + r\sin^{2} \alpha\right)\Bigg],
 \label{growth_rate_all}
\end{eqnarray}
where $r \equiv 1 - (u_{\textrm{\emph{th}}}/u_ *)^2$ quantifies the transport in the direction transverse to the flow (equations (\ref{slope_effect_direction}), (\ref{coeffay}) and (\ref{coeffby})). This function $\sigma$ is shown in figure~\ref{Sigma_unidirectional} in the plane $(\alpha,k)$ for the velocity ratios $u_{*}/u_{\textrm{\emph{th}}} = 1.6$ and $u_{*}/u_{\textrm{\emph{th}}} = 2.5$. The growth rate is positive (unstable modes) for small wavenumbers (large wavelengths) and transverse flows ($\alpha \to 0$). Large wavenumbers are stabilised by the flux relaxation process; longitudinal modes are stabilised by the slope effect on the transport direction. The marginal mode ($\sigma=0$) corresponds to the cutoff wavenumber
\begin{equation}
k_{\textrm{\emph{c}}}(\alpha) = \frac{1}{\cos \alpha \mathcal{A}_{0}}
\left[
\mathcal{B}_{0} -
\frac{\displaystyle\frac{u_{  th}}{u_{*}}\left(\displaystyle\frac{u_{  th}}{u_{*}}\cos^{2} \alpha + r\sin^{2} \alpha\right)}{\mu\left(\cos^{2} \alpha + \displaystyle\frac{r}{2}\sin^{2} \alpha\right)}
\right].
\label{cut_off_3D}
\end{equation}
The selected pattern is the one that maximises the growth rate, and we denote its wavenumber by $k_{\textrm{\emph{sat}}}$ and its orientation by $\alpha_{\textrm{\emph{sat}}}$. Taking the derivative of the growth rate \eqref{growth_rate} with respect to $k$ at constant $\alpha$, the most unstable wavenumber $k_\alpha$ for a given dune orientation satisfies
\begin{equation}
	k_\alpha = \frac{1}{\cos\alpha}\left(X^{1/3} - X^{-1/3}\right),
\end{equation}
where:
\begin{equation}
		X = \sqrt{1 + \left[\cos \alpha k_{\textrm{\emph{c}}}(\alpha)\right]^2} + \cos \alpha k_{\textrm{\emph{c}}}(\alpha).
\end{equation}
As the saturation length is significantly smaller than the dune wavelength, we can expand the above expressions in the limit $k_{{c}}(\alpha) \ll 1$, and we obtain:
\begin{equation}
	k_\alpha \sim \frac{2}{3}k_{{c}}(\alpha).
	\label{general_most_unstable}
\end{equation}
The overall unstable mode is then $k_{  max} = k_{\alpha_{  max}}$. Under a unidirectional flow, the selected orientation is perpendicular to the flow, i.e $\alpha_{  max} = 0^\circ$, which gives
\begin{equation}
k_{{max}} \sim \frac{2}{3}k_{\textrm{\emph{c}}}(0) = \frac{2}{3\mathcal{A}_{0}}
\left(
\mathcal{B}_{0} - \frac{1}{\mu}\frac{u_{  th}^{2}}{u_{*}^{2}}
\right),
\label{most_unstable2D}
\end{equation}
in agreement with the results of the two-dimensional linear stability analysis \citep{Andr02,Four10,Char13}.

Importantly, we see that the unstable mode exists only if the velocity ratio $u_{*}/u_{{th}}$ is larger than $1/(\sqrt{\mathcal{B}_{0}\mu})$. This condition is always satisfied with the above values of $\mathcal{B}_{0}$ and $\mu$ because $1/(\sqrt{\mathcal{B}_{0}\mu}) \simeq 0.8 < 1 \leq u_{*}/u_{{th}}$. Another related effect is that the selected wavelength $\lambda_{  max} = 2\pi/k_{{max}}$ tends to increase close to the threshold $u_{*}/u_{{th}} \to 1$. For infinitely large values of this velocity ratio, the slope effect on the transport direction vanishes and suppresses the stabilisation of longitudinal modes, so that $\sigma$ diverges in the double limit $k \to +\infty$ and $\alpha \to \pm 90^\circ$. Numerically, for the adopted values of $\mathcal{A}_{0}$, $\mathcal{B}_{0}$ and $\mu$, the regular shape of the growth rate, with a well-defined global maximum, is lost above $u_{*}/u_{{th}} \gtrsim 15$ (an unrealistic regime in practice).

\begin{figure}
	\centering
		\includegraphics[scale = 1]{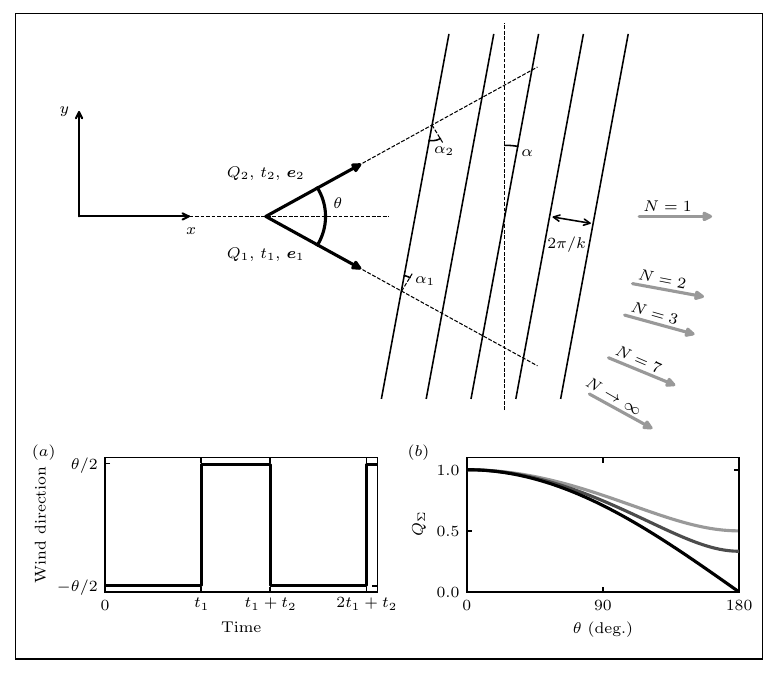}
	\caption{Characteristics of a bidirectional flow regime. The sketch shows the geometry of the system. The $x$-axis is defined as the bisector of the divergence angle $\theta$ between the two wind directions (black arrows, unit vectors $\boldsymbol{e}_{1,2}$). For the linear stability analysis, we consider a sinusoidal mode characterised by a wavenumber $k$ and an angle $\alpha$ with respect to the transverse direction ($y$-axis). The grey arrows are the directions of the resultant sand flux $\boldsymbol{Q}_\Sigma$ corresponding to different values of the transport ratio $N = (Q_1 t_1)/(Q_2 t_2)$.
Inset ($a$): alternation between the two wind directions with respect to time.
Inset ($b$): variation of the amplitude of the resultant flux $Q_\Sigma$ for $Q_1=Q_2$ as a function of $\theta$. From dark to light greys: increasing values of $N \in  \{1,\, 2,\, 3\}$.}
	\label{schema_bidi}
\end{figure}

\section{Dune growth in a bidirectional wind regime}
\label{bimodal}

We now extend the linear stability analysis developed above to investigate dune growth in multidirectional wind regimes. Here, we limit our analysis to bidirectional flow regimes, and compare our predictions to those of \citet{Rubi87} and \citet{Cour14} focused on dune orientation, and to the results of laboratory experiments presented in section~\ref{manip}.

\subsection{Linear contribution to the growth rate}
\label{bidi_th}

We consider two wind directions making an angle $\theta$, called the divergence angle, whose bisector defines the $x$-axis (figure~\ref{schema_bidi}). The wind blows alternately in directions of the unitary vectors $\boldsymbol{e}_{1}$ and $\boldsymbol{e}_{2}$ over durations $t_{1}$ and $t_{2}$, and with reference fluxes $Q_{1}$ and $Q_{2}$. In this bidirectional flow regime, we define the resultant flux $\boldsymbol{Q}_{\Sigma}$ as
\begin{equation}
\label{mean_sand_flux_vector}
	\boldsymbol{Q}_{\Sigma} = \frac{t_1}{t_1+t_2}Q_1 \boldsymbol{e}_{1} + \frac{t_2}{t_1+t_2}Q_2 \boldsymbol{e}_{2}.
\end{equation}
The intensity of this flux decreases with $\theta$ as the two winds compensate each other (figure~\ref{schema_bidi}$(b)$). We also define a time-averaged reference flux as
\begin{equation}
\label{mean_sand_flux}
	Q = \frac{t_1}{t_1+t_2}Q_1 + \frac{t_2}{t_1+t_2}Q_2.
\end{equation}
The period of wind reorientation $t_1+t_2$ is assumed to be small in comparison to the characteristic growth time of the bed perturbation of the order of $(1/\sigma_{\textrm{\emph{sat}}})L_{\textrm{\emph{sat}}} ^2/Q$ \citep{Rubi90}. Thus, the flow changes direction many times while the dune is still in its linear growth regime, during which its orientation and wavelength remain constant. The growth rate $\sigma_{\Sigma}$ of a mode characterised by $(\alpha,k)$ can then be expressed as the sum of the contributions of each wind:
\begin{equation}
\label{Sigma_bidi}
	\sigma_{\Sigma} = \frac{1}{N+1}\bigg[ N \sigma(Q_1, \alpha_1,k) + \sigma(Q_2, \alpha_2,k) \bigg],
\end{equation}
where $N = (Q_{1}t_{1})/(Q_{2}t_{2})$ is the transport ratio between the two winds \citep{Rubi87, Rubi90, Reff10, Cour14}. The growth rates $\sigma(Q_{1,2},\alpha_{1,2},k)$ are computed from that derived under a unidirectional flow \eqref{growth_rate_all} using
\begin{equation}
\label{def_alpha12}
	\alpha_{1,2} = \left[(\alpha \pm \frac{\theta}{2} +90^\circ) \mod 180^\circ \right] - 90^\circ,
\end{equation}
which are the angles between the pattern and the perpendicular to each flow direction (figures~\ref{schema} and \ref{schema_bidi}). For simplicity, we further assume that the winds blow with the same shear velocity $u_*$, so that $Q_1=Q_2=Q$. As a consequence, different values of $N$ correspond to different blowing times.

\begin{figure}
	\centering
		\includegraphics[width=\textwidth]{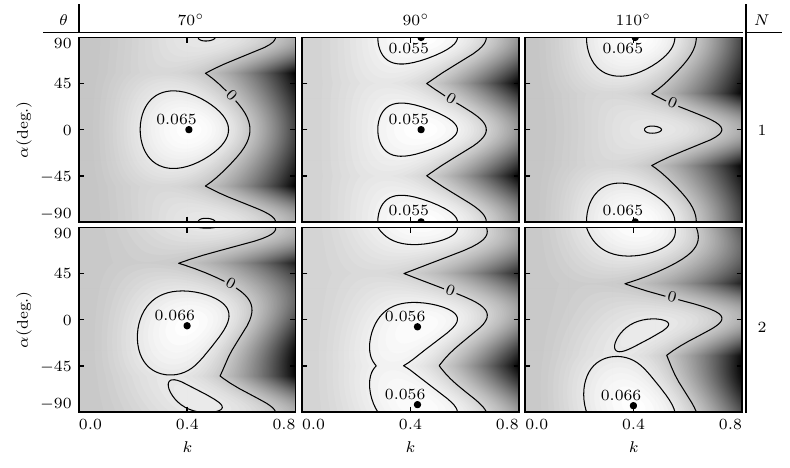}
	\caption{
Growth rate $\sigma_{\Sigma}$ as a function of the wavenumber $k$ and the orientation $\alpha$ of the bed perturbation for $u_*/u_{  th}=5$ and different values of the divergence angle $\theta$ (columns) and the transport ratio $N$ (rows). Black dots: most unstable mode $(\alpha_{\textrm{\emph{sat}}}, k_{\textrm{\emph{sat}}})$.}
	\label{sigma_bidi_maps}
\end{figure}

The growth rate $\sigma_{\Sigma}$ and the location of its maximum $(\alpha_{\textrm{\emph{sat}}}, k_{\textrm{\emph{sat}}})$ are shown in figure~\ref{sigma_bidi_maps} for some values of $N$ and $\theta$. This maximum integrates the effect of the two winds and gives the orientation and wavenumber of the emerging dune pattern. These selected dune orientations and wavelengths are the main output of the analysis. A third output is the propagation velocity of the selected mode. In the next subsections, we quantitatively discuss the behaviour of these selected quantities with respect to $N$, $\theta$ and $u_{*}/u_{{th}}$.

\subsection{Selection of the pattern orientation}
\label{transition_orientation}

\begin{figure}
	\centering
	\includegraphics[width=\textwidth]{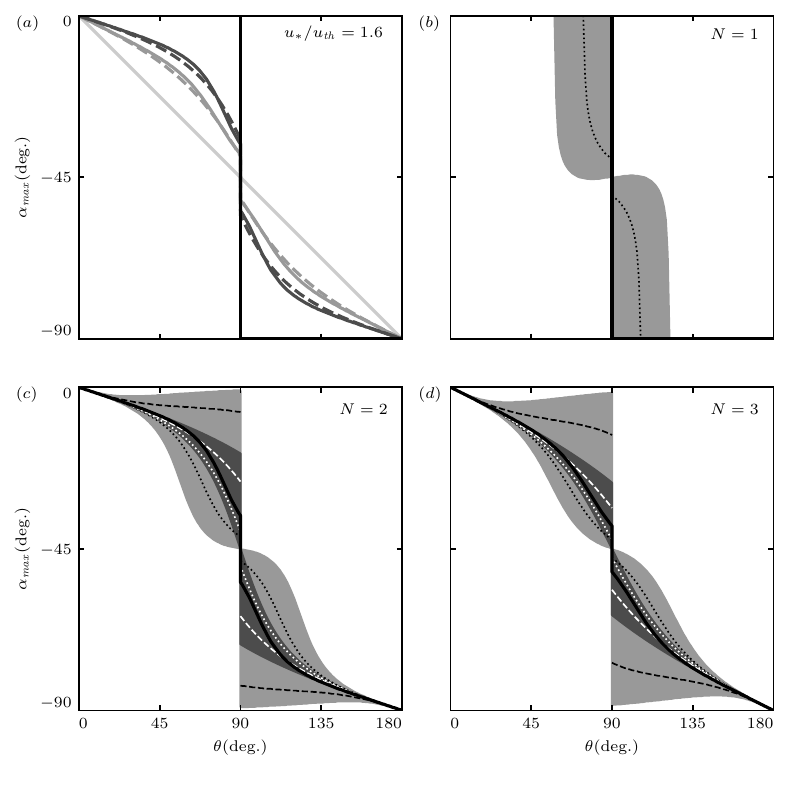}
	\caption{Pattern orientation $\alpha_{\textrm{\emph{sat}}}$ as a function of the divergence angle $\theta$ for various transport $N$ and velocity $u_*/u_{\textrm{\emph{th}}}$ ratios.
($a$) Solid lines: predictions of the linear stability analysis for $u_*/u_{\textrm{\emph{th}}} = 1.6$. Dashed lines: predictions of \citet{Cour14} for $\gamma = 1.7$. From dark to light greys: increasing values of $N \in  \{1,\, 2,\, 3,\, +\infty\}$.
($b$-$d$) Solid black lines: reference curves shown in $(a)$ for $u_*/u_{\textrm{\emph{th}}} = 1.6$. Dashed and dotted black lines: predictions of the linear stability analysis for $u_*/u_{\textrm{\emph{th}}} = 5$ and $u_*/u_{\textrm{\emph{th}}} = 1.2$, respectively. Dashed and dotted white lines: predictions of \citet{Cour14} for $\gamma = 0.5$ and $\gamma = 5$, respectively. Light grey region: all possible orientations predicted by this analysis when $u_*/u_{\textrm{\emph{th}}}$ is increased from $1.005$ to $15$. Dark grey region: all possible orientations predicted by \citet{Cour14} when $\gamma$ is varied from $0$ to $+\infty$ \eqref{Courrech_law}. Note that, in the strict case of $N=1$, the growth rate has an equal maximum for $(-\alpha_{\textrm{\emph{sat}}},k_{\textrm{\emph{sat}}})$.}
	\label{orientation_variation}
\end{figure}

As shown in figure~\ref{orientation_variation}, the orientation of the most unstable mode changes from transverse ($\alpha_{\textrm{\emph{sat}}} \approx 0^\circ$) when the divergence angle $\theta$ is small, to longitudinal ($\alpha_{\textrm{\emph{sat}}} \approx -90^\circ$) when the winds tends to be opposite (large divergence angle). This transition is the signature of a change in the normal to the crest fluxes \citep{Rubi87, Cour14}. When the divergence angle is smaller than $90^\circ$, the crests are more likely to be perpendicular to the resulting flux, as the sand is mainly transported along the $x$-direction. By contrast, when the divergence angle is larger than $90^\circ$, the crests are more likely to be aligned with the resulting flow as the sand is mostly transported along the $y$-direction. As discussed below, this transition depends on the transport ratio $N$ and the velocity ratio $u_*/u_{\textrm{\emph{th}}}$.

Let us first consider a fixed wind strength ($u_*/u_{\textrm{\emph{th}}} = 1.6$ in figure~\ref{orientation_variation}$(a)$) and discuss the effect of $N$. For winds of the same transport capacity ($N=1$), the switch occurs at the transition angle $\theta = 90^\circ$, with a discontinuous jump from a pattern perpendicular to the resultant flux direction ($\alpha_{\textrm{\emph{sat}}}=0^\circ$) to a pattern aligned with the resultant flux direction ($\alpha_{\textrm{\emph{sat}}}=-90^\circ$). This transition from a purely transverse to a purely longitudinal dune orientation is specific to the case $N=1$. When one of the two winds transports more than the other ($N>1$), intermediate orientations (oblique patterns) more perpendicular to the prevailing wind are permitted. There is still a discontinuous transition at the transition angle, but the jump decreases in amplitude as $N$ increases. This discontinuity is associated with two distinct maxima of equal growth rates (figure~\ref{sigma_bidi_maps}, central column). In the limit of unidirectional flows ($N \to +\infty$), the variation of $\alpha_{\textrm{\emph{sat}}}$ tends to a continuous linear function, which simply corresponds to the orientation perpendicular to the dominant wind.

The effect of the velocity ratio $u_*/u_{\textrm{\emph{th}}}$, displayed in figure~\ref{orientation_variation} with light grey regions, can be summarised as follows. Strong winds tend to enhance the orientation discontinuity at $\theta = 90^\circ$, keeping $\alpha_{\textrm{\emph{sat}}}$ either closer to $0^\circ$ ($\theta < 90^\circ$) or to $-90^\circ$ ($\theta > 90^\circ$). This corresponds to a selected orientation less perpendicular to the prevailing wind. By contrast, winds close to the threshold induce a more continuous change in orientation, i.e. closer to the perpendicular to the prevailing wind. When $u_*/u_{\textrm{\emph{th}}} \gg 1$, slope effects become negligible and oblique modes stabilised by transverse transport are less damped for large velocity ratios $u_*/u_{\textrm{\emph{th}}}$ (figure~\ref{Sigma_unidirectional} in the case of a unidirectional wind). When the growth rates associated with the two winds are combined, the `cost' to remain away from the perpendicular to the prevailing wind is not so strong, and the selected orientation reflects this trend. When $u_*/u_{\textrm{\emph{th}}} \to 1$, slope effects are strong, and mainly affect longitudinal transport, i.e. the coefficient $b_x$ \eqref{coeffbx}, where two terms compete. The first one, of hydrodynamic origin, scales with $\cos^2\alpha_{1,2}$, as deduced from the properties of the potential flow. The second one is related to the variation of the threshold with the slope, i.e. proportional to $-\cos\alpha_{1,2}$. This difference in the dependence with $\alpha_{1,2}$ favours a crest orientation perpendicular to the prevailing wind, for which the destabilising hydrodynamics is the strongest.

\subsection{Comparison with the dimensional analysis of \citet{Cour14}}
\label{ModelCourrech}

The prediction of the dune orientation by this linear stability analysis can be directly compared to that of \citet{Cour14}. Based on an estimation of the sand flux at the dune crest, these authors computed by dimensional analysis the dune growth rate accounting for the wind speed-up by the topography (appendix \ref{Deriv_cour}). The selected pattern orientation is then such that it maximises
\begin{equation}
	\label{Courrech_law}
	\sigma_{\Sigma} \propto N\cos\alpha_1 + N\gamma\cos^2 \alpha_1 + \cos\alpha_2 + \gamma\cos^2 \alpha_2 ,
\end{equation}
where $\gamma$ quantifies the relative increase of sand flux between the flat bed and the dune crest. For $\gamma = 1.7$ (a value also derived from field observations and numerical models by \citet{Gao15}), the results of that model are found to quantitatively match our predictions for $u_*/u_{\textrm{\emph{th}}} = 1.6$ (see figure~\ref{orientation_variation}). When $\gamma$ is varied, the curve $\alpha_{\textrm{\emph{sat}}}(\theta)$ changes, and the dark grey area in figure~\ref{orientation_variation}$(b-d)$ displays the whole range of possible orientations, showing an overall similar behaviour.

The present linear analysis corresponds to the limit $k\zeta \ll 1$, whereas the computation of \citet{Cour14} considers a dune with a finite aspect ratio, for which nonlinearities are expected. In a similar fashion to \citet{Cour15} and \citet{Gao15} (in Supplementary Material), the parameter $\gamma$ can be related to the parameters of the present model as
\begin{equation}
\label{gamma}
	\gamma = \frac{|\hat{q}_{  sat}|}{q_{  sat}} = \frac{k\zeta}{r} \sqrt{\mathcal{A}_0^{2} + \mathcal{B}_0^{2}},
\end{equation}
where $\hat{q}_{  sat}$ is the Fourier transform of the perturbation of the saturated flux. With the values $\mathcal{A}_0=3.5$, $\mathcal{B}_0=2$ and $u_*/u_{  th} = 1.6$ (or equivalently $r \simeq 0.7$, see \eqref{growth_rate_all}), the parameter $\gamma = 1.7$ corresponds to a bed perturbation of aspect ratio of $2\zeta/\lambda = 1/12$. This value is consistent with the typical aspect ratio of fully developed dunes \citep{Part06, Badd07, Elbe08}. This general consistency between the two approaches emphasises the role played by the normal-to-the-crest components of transport in the selection of dune orientation during the linear but also nonlinear phase of dune growth.

\begin{figure}
	\centering
	\includegraphics[width=\textwidth]{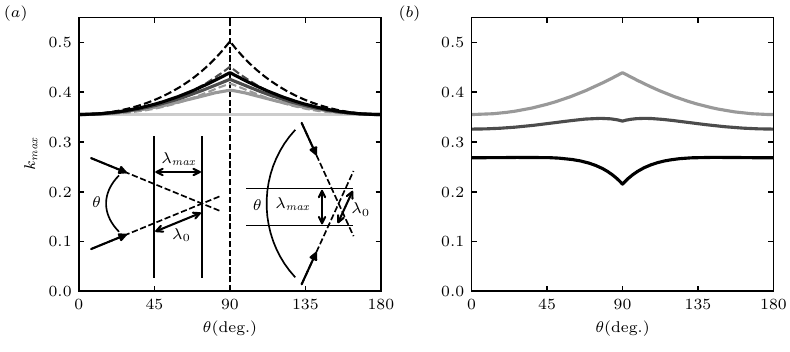}
	\label{orientation_comparaison}
	\caption{Selected wavenumber $k_{\textrm{\emph{sat}}}$ as a function of the divergence angle $\theta$ for various transport $N$ and velocity $u_*/u_{\textrm{\emph{th}}}$ ratios. As for comparison with figure~\ref{Wavelength_exp}, note that for $\lambda_{\textrm{\emph{sat}}} = 2\pi/k_{\textrm{\emph{sat}}}$, the increasing/decreasing trends are opposite.
$(a)$ Solid lines: prediction of the linear stability analysis for $u_*/u_{\textrm{\emph{th}}} = 5$. Dashed lines: prediction using the geometric scalings, sketched in the insets for $N=1$ (\ref{lambda_geometric_lessthan90}, \ref{lambda_geometric_morethan90}). From dark to light greys: increasing values of $N \in  \{1,\, 2,\, 3,\, +\infty\}$.
$(b)$ Prediction of the linear stability analysis for $N=1$. From dark to light greys: increasing values of $u_*/u_{\textrm{\emph{th}}} \in  \{1.6,\, 2.5,\, 5\}$.}
	\label{wavenumber_variation}
\end{figure}

\subsection{Selection of the pattern wavenumber}
\label{pattern_wavenumber}

The second key output of the linear stability analysis is the selected pattern wavenumber $k_{\textrm{\emph{sat}}}$. Remember that $k$ is here made dimensionless by $L_{\textrm{\emph{sat}}} $, the only characteristic length scale of the problem. The shape of $k_{\textrm{\emph{sat}}}$ as a function of the divergence angle $\theta$ is displayed in figure~\ref{wavenumber_variation}, for different values of the transport $N$ and velocity $u_*/u_{\textrm{\emph{th}}}$ ratios. We can see a non-monotonic behaviour, with variations that can reach $\simeq 30\%$ close to the transition angle $\theta = 90 ^\circ$. Interestingly, these variations can be positive or negative, depending on whether the wind strength is far from or close to the transport threshold.

For symmetric winds ($N=1$) and in the limit of large shear velocities, a geometric argument sketched in the insets of figure~\ref{wavenumber_variation}$(a)$ can explain the dependence of the wavelength on the divergence angle (solid curves in that figure). Let us call $\lambda_0$ the selected wavelength in the unidirectional regime, and $k_0=2\pi/\lambda_0$ the corresponding wavenumber. Far from the transport threshold, and thus without any slope effect, the balance between stabilising and destabilising processes occurs along the wind direction. In a bidirectional wind regime, the selected orientation is oblique to each flow direction, and we can infer that the growth rate is maximum when the wavelength $\lambda_0$ is selected in the direction of each wind.

This argument can be made more precise as follows. Far from the transport threshold, the second term in \eqref{cut_off_3D} associated with the slope effects vanishes asymptotically as $u_{  th}/u_* \to 0$, and this leads to
\begin{equation}
	k_\alpha\sim \frac{2}{3}k_c(\alpha) \sim \frac{2}{3}\frac{1}{\cos\alpha}\frac{\mathcal{A}_{0}}{\mathcal{B}_{0}} = \frac{1}{\cos\alpha}k_0.
	\label{vent_fort}
\end{equation}
Observing that the most unstable mode maximises the growth rate associated with each of the two winds, we obtain
\begin{equation}
k_{  max} = k_{\alpha_1(\alpha_{  max})} = k_{\alpha_2(\alpha_{  max})}
\end{equation}
implying that $\cos\alpha_{1}(\alpha_{  max}) = \cos\alpha_{2}(\alpha_{  max})$.
The corresponding wavelength is
\begin{equation}
	\lambda_{  max} \sim \cos\alpha_{1,2}(\alpha_{  max})\lambda_0.
\end{equation}
In particular, for a symmetrical wind regime ($N = 1$) and using \eqref{def_alpha12}, we have
\begin{eqnarray}
	\cos\alpha_{1,2}(\alpha_{\textrm{\emph{sat}}}) = \cos \frac{\theta}{2} \quad& \textrm{for} \quad \theta \le 90^\circ, \\
	\cos\alpha_{1,2}(\alpha_{\textrm{\emph{sat}}}) = \sin \frac{\theta}{2} \quad& \textrm{for} \quad \theta \ge 90^\circ,
\end{eqnarray}
which gives
\begin{eqnarray}
\lambda_{  max} = \lambda_0 \cos \frac{\theta}{2} \quad \textrm{or} \quad k_{  max} = k_0 / \cos \frac{\theta}{2} \quad& \textrm{for} \quad \theta \le 90^\circ,
\label{lambda_geometric_lessthan90} \\
 \lambda_{  max} = \lambda_0 \sin \frac{\theta}{2} \quad \textrm{or} \quad k_{  max} = k_0 / \sin \frac{\theta}{2} \quad& \textrm{for} \quad \theta \ge 90^\circ.
\label{lambda_geometric_morethan90}
\end{eqnarray}
For $N>1$, we do not have an analytic expression for $\cos\alpha_{1,2}(\alpha_{  max})$ as a function of $\theta$, but this can be done numerically. These asymptotic predictions of the wavenumber are shown for different values of the transport ratio $N$ in figure~\ref{wavenumber_variation}$(a)$ in dotted lines. We recover with this geometric argument the trends predicted by the linear stability analysis at finite $u_*/u_{\textrm{\emph{th}}}$, and in particular the flattening of the curve for larger values of $N$. We notice, however, a slight overestimation of the peak at $\theta = 90^\circ$, because the slope effects are still present.

The role of the velocity ratio $u_*/u_{\textrm{\emph{th}}}$ on the selected wavenumber is shown in figure~\ref{wavenumber_variation}$(b)$ for $N=1$. When the wind strength decreases, the threshold and direction of transport are changed as the slope effects are increasingly important \eqref{slope_effect_direction}. First, because the amplitude of the hydrodynamic shift is reduced (the `$\mathcal{B}$ effect'), the selected wavenumber is smaller, as in the unidirectional case \eqref{most_unstable2D}. Moreover, for a bidirectional flow regime, $k_{\textrm{\emph{sat}}}$ is even smaller near the transition angle $\theta = 90^\circ$. This is a consequence of the stabilising action of the transverse component of sediment transport, as discussed in section~\ref{sed_trans}. This effect is the strongest at the transition angle, as it depends on the component of the slope perpendicular to the flow direction. As a result, the curves close to and far from the transport threshold show opposite behaviours. For intermediate wind strengths, the curve $k_{\textrm{\emph{sat}}} (\theta)$ is almost flat, as the geometric and slope mechanisms compensate each other.

\begin{figure}
	\centering
	\includegraphics[width=\textwidth]{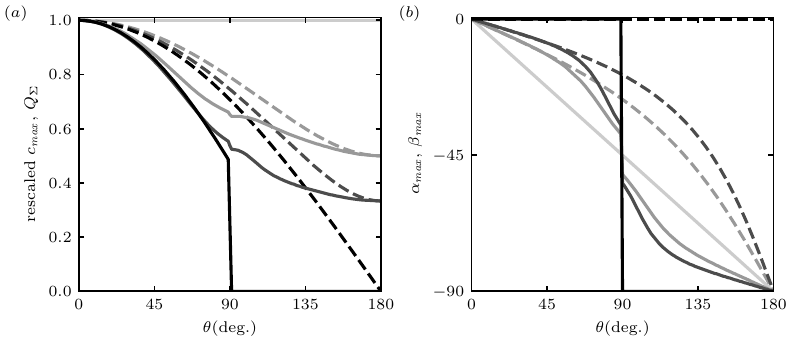}
	\caption{Propagation velocity of the most unstable mode $\boldsymbol{c}_\Sigma(\alpha_{\textrm{\emph{sat}}},k_{\textrm{\emph{sat}}})$ and resultant flux $\boldsymbol{Q}_\Sigma$ as functions of the divergence angle $\theta$ for $u_*/u_{\textrm{\emph{th}}} = 1.6$ and various transport ratio $N$. From dark to light greys: increasing values of $N \in  \{1,\, 2,\, 3,\, +\infty\}$.
$(a)$ Solid lines: propagation velocity $c_{\textrm{\emph{sat}}}$ of the most unstable mode. Dashed lines: resultant flux $Q_\Sigma$. Both are rescaled by their value in $\theta = 0 ^ \circ$.
$(b)$ Solid lines: Direction of the resultant flux $\beta_{\textrm{\emph{sat}}}$ measured with respect to the
$x$-axis. For comparison, the pattern orientation $\alpha_{\textrm{\emph{sat}}}$ (see figure~\ref{orientation_variation}$(a)$) is also plotted (dashed lines).
}
	\label{celerity_fig}
\end{figure}

\subsection{Propagation velocity of the most unstable mode}
\label{pattern_velocity}

Similarly to the growth rate \eqref{Sigma_bidi}, the propagation velocity of the emerging pattern can be expressed as the sum of the contribution of each wind:
\begin{equation}
\label{Cel_bidi}
	\boldsymbol{c}_\Sigma = \frac{1}{N+1}\bigg[ N c(Q_1,\alpha_1,k) + \textrm{sgn}(\theta - 90)c(Q_2,\alpha_2,k) \bigg]\frac{\boldsymbol{k}}{k}.
\end{equation}
When the divergence angle is larger than $90^\circ$, the displacements of the pattern induced by each of the winds are opposite, resulting in the term $\textrm{sgn}(\theta - 90)$. From the propagation velocity can be computed that of the most unstable mode $c_{\textrm{\emph{sat}}} = \| \boldsymbol{c}_\Sigma(\alpha_{\textrm{\emph{sat}}},k_{\textrm{\emph{sat}}}) \|$. This quantity, as well as the resultant flux $Q_\Sigma$ on a flat bed \eqref{mean_sand_flux_vector} and its direction quantified by $\beta_{\textrm{\emph{sat}}} = \angle (\boldsymbol{Q}_\Sigma, \boldsymbol{e}_x)$, are shown in figure~\ref{celerity_fig} as functions of the divergence angle $\theta$ for a velocity ratio $u_*/u_{\textrm{\emph{th}}} = 1.6$ and different values of the transport ratio $N$.

The propagation velocity is controlled by the component of $Q_\Sigma$ perpendicular to the crest and by the pattern wavelength. For $\theta < 70 ^ \circ$, the overall behaviour is simple: $c_{\textrm{\emph{sat}}}$ and $Q_\Sigma$ vary in the same way. Their directions are almost the same, as the flux direction is not significantly modified by the pattern. As shown in figure \ref{celerity_fig}$(a)$ by the dashed lines, the resultant flux and thus the propagation velocity decrease for wider divergence angles, because the winds increasingly compensate each other. This compensation is weaker when one wind is stronger than the other ($N > 1$). For $\theta > 110 ^ \circ$, the resultant sand flux is mostly aligned with the dune crest leading to a slower propagation. The velocity even vanishes for $N = 1$. Finally, around the transition angle $\theta = 90 ^ \circ$, the resultant flux direction is strongly modified by the pattern orientation. In addition, the wavelength is also affected by the slope effects. Both effects lead to a complex behaviour of the propagation velocity, which depends on the velocity ratio $u_*/u_{\textrm{\emph{th}}}$.

\section{Subaqueous experiments}
\label{manip}

Subaqueous experiments are another test of the proposed linear stability analysis, especially with regard to the selected pattern wavelength. In laboratory experiments, an initial flat sand bed is subjected to a bidirectional  flow  regime leading to the formation of ripples. Although subaqueous bedload differs from aeolian saltation in several aspects, these ripples originate from the same physical mechanism as dunes, i.e. a destabilising role of the hydrodynamics balanced by a stabilising role of transport relaxation \citep{Char13}. One can then take these ripples as analogous to aeolian dunes, with downscaled size and time scales \citep{Hers02, Reff10, Cour14}. For that reason, and to be consistent with the aeolian context of this work, we shall use the term \emph{dunes} for these subaqueous bedforms in the rest of this section.

\begin{figure}
	\includegraphics[width = \textwidth]{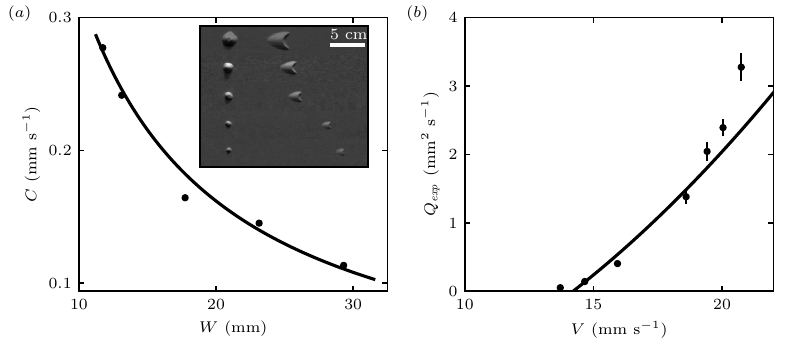}
	\caption{Estimation of the transport threshold in the experimental set-up.
$(a)$ Velocity $C$ of the barchans as a function of their widths $W$ for a plate velocity $V = 20.8~\textrm{cm}~\textrm{s}^{-1}$. The black dots are the experimental data, and the solid line is the fit of the Bagnold-like law \eqref{bagnold_relation_exp} with $W_0 = 0.1 \pm 0.2~\textrm{mm}$ and $Q_{\textrm{\emph{exp}}}  = 3.3 \pm 0.2~\textrm{mm}^2~\textrm{s}^{-1}$. The inset is a photomontage showing the corresponding initial sand piles and subsequent dunes after $169$~blows.
$(b)$ Transport law with the effective flux $Q_{\textrm{\emph{exp}}} $ extracted from the barchan velocities for different values of $V$. Error bars correspond to $95\%$-confidence intervals in the fitting process. The solid line shows the fit of $Q_{\textrm{\emph{exp}}}  \propto (V^2 - V_{\textrm{\emph{th}}}^2)$, which gives $V_{\textrm{\emph{th}}} = 14.2 \pm 0.1~\textrm{cm}~\textrm{s}^{-1}$.}
	\label{Exp_transportlaw}
\end{figure}

\subsection{Experimental set-up and calibration}

The set-up is the one used by \citet{Reff10} and \citet{Cour14}, which we briefly recall here. The granular material (ceramic beads of diameter $d=90~\mu\textrm{m}$ and density $\rho = 3800~\textrm{kg}~\textrm{m}^{-3}$) lies on a $80~\textrm{cm}$ wide square plate, which can move from one side to the other of a $2~\textrm{m}$ long and $1~\textrm{m}$ wide water tank. The tank is filled to a water height of $30~\textrm{cm}$ and the plate stands at mid-height. The plate first quickly moves in one direction at a velocity that generates sediment transport, and slowly translates in the opposite way at $2~\textrm{cm}~\textrm{s}^{-1}$ to prevent grain motion. This sequence, which we call a blow, simulates a unidirectional flow over the bed. At the centre of the plate, a $70~\textrm{cm}$ diameter disc can rotate to change the relative direction of the flow with respect to the bed. A camera set above the plate allows us to follow the evolution of the bed after each blow.

The quick motion of the plate is decomposed into two steps. The plate accelerates from rest to the final velocity in $250~\textrm{ms}$ and then gently decelerates in approximately $4~\textrm{s}$ to prevent a backflow. For simplicity, we choose the mean plate speed $V$ as the parameter representing the plate movement. Because the translation of the plate puts the fluid into motion inside a boundary layer that grows over time, the shear stress on the bed is maximum during the acceleration stage and most of the transport takes place during this first phase. The time needed for the boundary layer to reach the water surface is much larger than the translation time, so that the flow can be considered as unbounded.

The transport threshold was not quantitatively determined for this experimental set-up in previous studies. Because our predictions depend on the velocity ratio $u_*/u_{{th}}$, it is important to know how far from the threshold experiments are run. We can provide a first estimation with the observation that no pattern is visible after $2020~\textrm{blows}$ for a mean plate velocity $V=13.7~\textrm{cm}~\textrm{s}^{-1}$. However, since growth rates are much smaller close to the transport threshold, this estimation is made more precise by using the migration velocity $C$ of barchan dunes of different sizes submitted to a unidirectional flow regime. We vary the mean plate translation velocity $V$ while keeping the plate acceleration time constant. Figure \ref{Exp_transportlaw}$(a)$ shows the velocity of barchans as a function of their width for a fixed translation motion ($V=20.8~\textrm{cm}~\textrm{s}^{-1}$). The barchan velocity is smaller for larger barchans, and is well fitted by the Bagnold-like law \citep{Bagn41}
\begin{equation}
\label{bagnold_relation_exp}
C = \frac{Q_{\textrm{\emph{exp}}} }{W+W_0}.
\end{equation}
This propagation law, usually expressed in terms of the dune height $H$, derives from the conservation of mass assuming a constant aspect ratio for the dunes. Under this assumption, the height, the spanwise width $W$ and the streamwise length of barchans are proportional to each other, and $W$ is a more convenient observable on the photos. In \eqref{bagnold_relation_exp} $W_0$ accounts for a correction of Bagnold's law at small sizes \citep{Elbe08}. As illustrated in figure \ref{Exp_transportlaw}$(a)$, this law agrees well with the velocity measurements, and its fit allows us to extract an effective flux $Q_{\textrm{\emph{exp}}} $ from the different datasets corresponding to various values of $V$. In this fitting procedure, the adjustable parameter $W_0$ has been constrained to be the same for all the data.

Plotting the effective flux $Q_{\textrm{\emph{exp}}} $ as a function of the plate velocity (figure~\ref{Exp_transportlaw}$(b)$), we see as expected an increasing behaviour with a threshold below which the barchans do not move. Although the sediment transport is intermittent in this experiment, we fit the data with the usual transport law $Q_{\textrm{\emph{exp}}}  \propto (V^2 - V_{\textrm{\emph{th}}}^2)$. This gives a threshold plate velocity $V_{\textrm{\emph{th}}} = 14.2 \pm 0.1~\textrm{cm}~\textrm{s}^{-1}$. Note that this value does not depend much on the precise choice of the fitting function.

\begin{figure}
	\includegraphics[width = \textwidth]{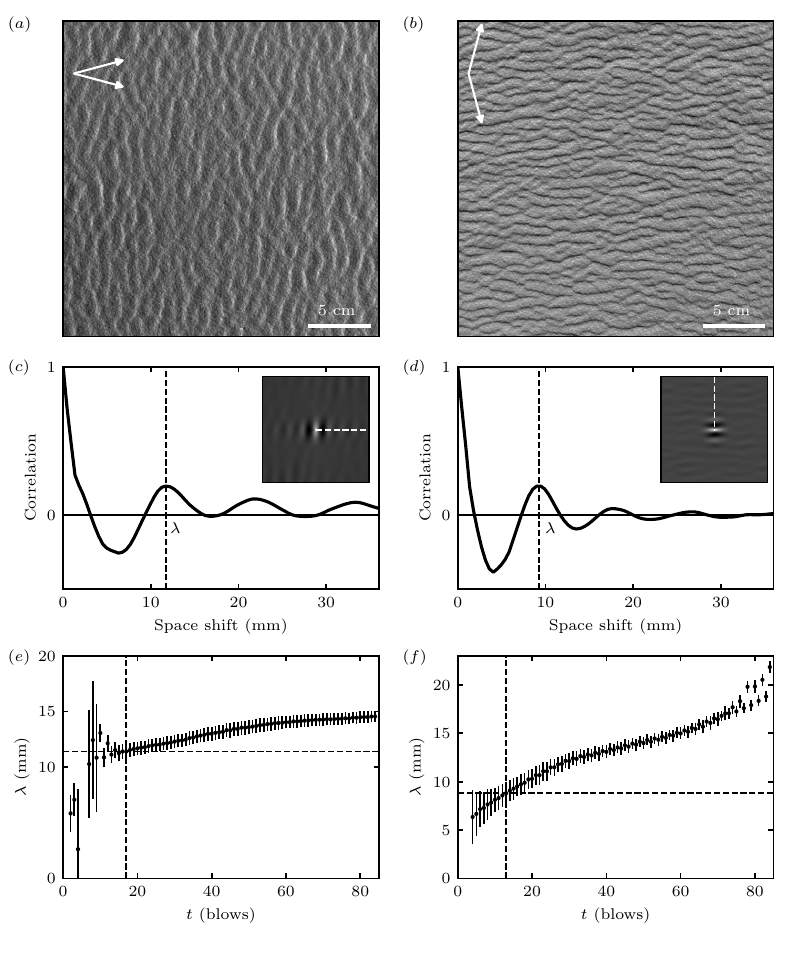}
	\caption{Pattern formation from a flat bed in a bidirectional flow regime for $\theta = 30^\circ$ (left) and $\theta = 150^\circ$ (right), for $V/V_{\textrm{\emph{th}}} = 1.4$.
($a,b$) pictures after 17 and 13 blows respectively, starting from a flat bed.
($c,d$) autocorrelation of the above pictures. Insets: central zone of the autocorrelation matrix encoded in grey scales, which gives the pattern orientation and wavelength. The black curves are the autocorrelation signal along the white dashed lines, i.e. perpendicular to the dune orientation. The measured wavelength $\lambda$ corresponds to the position of the first maximum of the autocorrelation curve.
($e,f$) wavelength as a function of time. Dotted lines: initial wavelength and corresponding time at which the spatial coherence of the autocorrelation signal extends over one wavelength. Error bars encode the quality of this signal.}
	\label{Image_analysis}
\end{figure}

\subsection{Experimental results}

We present here results in the case of symmetric winds ($N=1$): the central disc of the plate is rotated by $\theta$ and $-\theta$ alternately after every blow. Starting from an initial flat bed, dunes rapidly form after a few blows with a characteristic wavelength and orientation, which are measured by picture autocorrelation (see figure~\ref{Image_analysis}). We mainly focus below on the dependence of the wavelength with the divergence angle $\theta$, as the behaviour of the pattern orientation in this experimental set-up has already been discussed in \citet{Cour14}, showing agreement with their approach (see section~\ref{ModelCourrech}), and thus the present analysis (figure~\ref{orientation_variation}).

Figure~\ref{Image_analysis} displays results of two well-contrasted cases with a small and a large divergence angle ($\theta=30^\circ$ and $\theta=150^\circ$, respectively), keeping the same plate velocity ratio $V/V_{{th}}=1.4$. When $\theta = 30^\circ$, transverse dunes form perpendicular to the resultant transport direction with an initial wavelength of $11.8 \pm 0.7~\textrm{mm}$. When $\theta = 150^\circ$, longitudinal dunes form parallel to the resultant transport direction with an initial wavelength of $8.8 \pm 0.5~\textrm{mm}$. After the emergence of the pattern, some coarsening is observed, with an increase of the wavelength with time.

\begin{figure}
	\includegraphics[width = \textwidth]{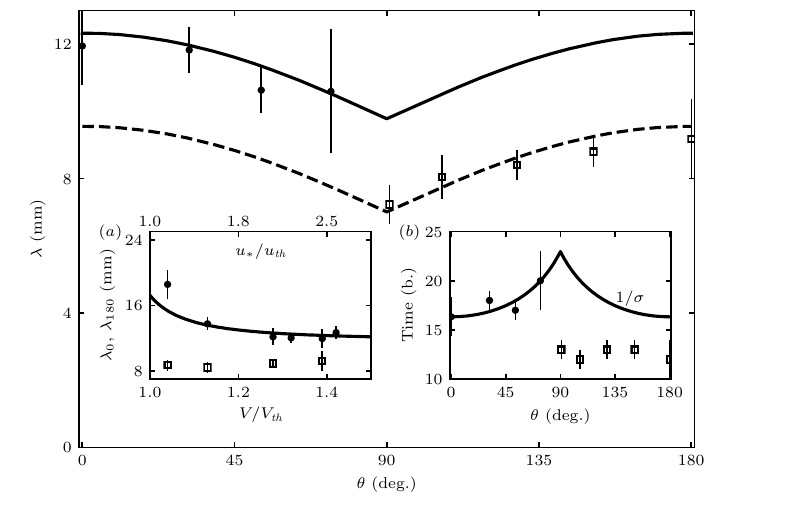}
	\caption{Initial wavelength as a function of the divergence angle $\theta$ (symmetric winds $N=1$). Symbols: experimental data for $V/V_{\textrm{\emph{th}}} = 1.4$. Solid line: model prediction for $u_*/u_{\textrm{\emph{th}}} = 2.5$ and $L_{\textrm{\emph{sat}}} = 1.1~\textrm{mm}$. Dashed line: solid line shifted below and adjusted in $\theta = 180^\circ$ (we subtracted $(\lambda_0 - \lambda_{180})$).
Inset $(a)$: initial wavelength as a function of the rescaled plate velocity for $\theta = 0^\circ$ ($\lambda_0$, bullets) and $\theta = 180^\circ$ ($\lambda_{180}$, squares). Solid line: fit of the linear stability analysis under a unidirectional flow.
It gives $L_{  sat} = 1.1 \pm 0.1$~mm and $C_v = 4 \pm 2$, see~\eqref{Fit_2D}.
Inset $(b)$: Characteristic time (in blows) of pattern emergence as a function of $\theta$. Symbols: experimental data for $V/V_{\textrm{\emph{th}}} = 1.4$. Solid line: model prediction for $u_*/u_{\textrm{\emph{th}}} = 2.5$, adjusted in $\theta = 0^\circ$. For all model predictions, $\mathcal{A}_0 =4$ and $\mathcal{B}_0 =4$ (see text).}
	\label{Wavelength_exp}
\end{figure}

Figure~\ref{Wavelength_exp} shows the behaviour of the wavelength when the divergence angle $\theta$ is varied over the whole range. When $\theta < 90^{\circ}$, transverse dunes form. Their wavelength is maximum for $\theta=0^\circ$ and decreases when $\theta$ increases. When $\theta > 90^{\circ}$, longitudinal dunes form and exhibit opposite behaviour: their wavelength increases with $\theta$ and is maximum for $\theta=180^{\circ}$. Surprisingly, this variation of the dune wavelength with the divergence angle is not symmetric with respect to $\theta=90^{\circ}$, and longitudinal dunes systematically form at smaller wavelengths than transverse ones. This is true for all values of the plate velocity ratio $V/V_{{th}}$ that we have tested (see figure~\ref{Wavelength_exp}$(a)$). Furthermore, the wavelength under a unidirectional wind $\lambda_0$ is found to decrease with $V/V_{{th}}$. Wavelengths are significantly larger close to the transport threshold in comparison to the asymptotic value of approximately $12~\textrm{mm}$ corresponding to large $V/V_{{th}}$. By contrast, the wavelength $\lambda_{180}$ under two opposed flows ($\theta=180^\circ$) is approximately $9~\textrm{mm}$ and does not depend on the plate velocity.

\begin{figure}
	\includegraphics[width = \textwidth]{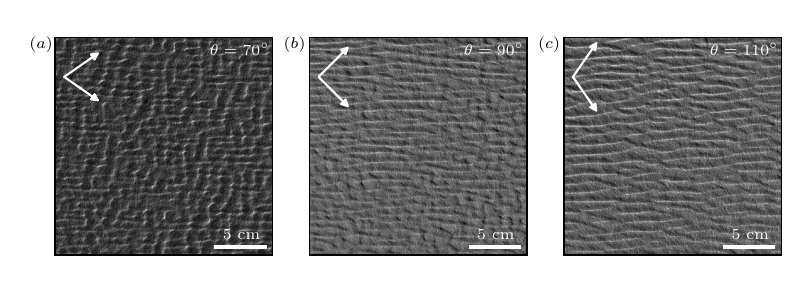}
	\caption{Photographs of the dune pattern in the experiment across the transition from transverse to longitudinal orientations, starting from a flat bed, for $V/V_{\textrm{\emph{th}}} = 1.4$. Flow directions are shown with arrows. From left to right, the pictures have been taken after $30$, $32$, and $30$ blows, respectively}
	\label{patterns_2modes}
\end{figure}

We can also estimate the characteristic time needed to observe incipient dunes in the experiment. This time is a proxy for the inverse of the instability growth rate. It is displayed in figure~\ref{Wavelength_exp}$(b)$ as a function of the divergence angle $\theta$. Both transverse and longitudinal dunes emerge typically after a dozen blows, but systematically more rapidly in the case of longitudinal ones. The pattern coarsening is also quicker for longitudinal dunes than for transverse dunes (figure~\ref{Image_analysis}$(e,f)$), an observation already reported by \citet{Reff10}. At the transition between the two dune orientations, close to $\theta=90^{\circ}$, the two modes are observed simultaneously in the building dune pattern, as shown in figure \ref{patterns_2modes}. The longitudinal orientation seems to overcome the transverse one at the transition. This asymmetry appears to be specific to incipient dunes, as the transition is sharper once dunes are more mature \citep{Reff10}.

\subsection{Comparison with the theoretical predictions}

In order to make comparisons between these experiments and the linear stability analysis developed for aeolian dunes, the model should be adapted to the subaqueous case. The transport law for bedload is different from that of saltation, with a power $3/2$ rather than the linear relationship \eqref{transport_law_tau} between the saturated sediment flux and the basal shear stress. However, the details of this transport law affect the linear stability analysis by modifying factors in $($\ref{coeffax} - \ref{coeffby}$)$ only, and we have checked that all the discussed results (e.g., increase of wavelength close to transport threshold, decrease of wavelength close to $\theta = 90^\circ$) are qualitatively robust with respect to any choice of transport law. Also, the transient nature of the flow and grain transport in the experiment prevents us from considering this configuration as a standard bedload situation, where we could quantitatively rely on well-calibrated transport laws. For these reasons, we opted for simplicity and chose to keep the linear transport law \eqref{transport_law_tau}.

Even if we do not necessarily aim at a quantitative matching of the model with the experimental data, the values for the basal shear stress coefficients can nevertheless be more consistent with the experimental case. In fact, the coefficients $\mathcal{A}_0=3.5$ and $\mathcal{B}_0 = 2$ are typical values measured in the field for aeolian dunes of $10~\textrm{m}$ wavelength \citep{Clau13}. These coefficients are weak (logarithmic) functions of the pattern wavenumber at small $k z_0$, but show stronger variations in the range $10^{-4} < kz_0 < 10^{-2}$ due to an anomalous hydrodynamic response of the flow to the bed perturbation, as first experimentally evidenced by \citet{Zilk77} and \citet{Fred88}. In the experiment, the pattern wavelength is of the order of $10$~mm, i.e. wavenumber $k=2\pi/\lambda \simeq 5.7 \times 10^{2}$~m$^{-1}$. The hydrodynamic roughness $z_0$ is more difficult to estimate under these transient water flow and sediment transport conditions. Taking it as a small fraction of the grain diameter, we typically have $k z_0 \simeq 10^{-3}$, i.e. two orders of magnitude larger than in the aeolian case. For such a value, the coefficients become $\mathcal{A}_0 = 4$ and $\mathcal{B}_0 = 4$ \citep{Char13}, and we take these values to be fixed for all theoretical computations in this section. Note that increasing these coefficients corresponds to a larger relative weight of hydrodynamic effects compared to slope effects. Interestingly, with these slightly larger values of $\mathcal{A}_0$ and $\mathcal{B}_0$ compared to the aeolian case, the slope effects are no longer strong enough to induce an increase of the wavelength at $\theta = 90^\circ$ close to the transport threshold (figure 7(b)). Only the hydrodynamic effect leading to the decrease of the wavelength at the transition angle remains. This is consistent with the experimental observations.

The most unstable wavelength $\lambda_{  max}$ selected by the linear stability analysis depends on the velocity ratio $u_*/u_{\textrm{\emph{th}}}$. This ratio is not directly equal to the plate velocity ratio $V/V_{\textrm{\emph{th}}}$ in the experiment because of the transient nature of the flow. Here, we assume that they are linearly related to each other :
\begin{equation}
		\left(\frac{V}{V_{\textrm{\emph{th}}}} - 1\right) = C_v \left(\frac{u_*}{u_{\textrm{\emph{th}}}} - 1\right),
	\label{Fit_2D}
\end{equation}
where $C_v $ is an adjustable constant. In the fitting procedure, the most unstable wavelength is thus a function of $V/V_{\textrm{\emph{th}}}$ with parameters $C_v $ and $L_{  sat}$. For a unidirectional flow ($\theta = 0^\circ$), the comparison between data and theory is shown in figure~\ref{Wavelength_exp}$(a)$. The observed increase of the wavelength close to the transport threshold is quantitatively reproduced by the model. This effect is due to the sensitivity of the threshold to the bed slope. The fit gives $L_{  sat} = 1.1 \pm 0.1$~mm (i.e. $\simeq 12$ grain diameters) and $C_v = 4 \pm 2$. These values are now fixed and used for the subsequent computation of theoretical curves at all divergence angles, without any further adjustment. In particular, we run all experiments at $V/V_{\textrm{\emph{th}}} = 1.4$, which corresponds to a velocity ratio $u_*/u_{\textrm{\emph{th}}} = 2.5$, from \eqref{Fit_2D}.

Figure~\ref{Wavelength_exp} shows that the linear stability analysis captures well the variation of the wavelength for divergence angles smaller than $90^\circ$ (solid line). For divergence angles larger than $90^\circ$, the experimental data are significantly smaller than predicted whereas the model predicts a symmetric variation of the wavelength with respect to the transition angle $\theta = 90^\circ$. Nevertheless, the decrease of the wavelength close to  $\theta = 90^\circ$ is recovered relative to the wavelength at $\theta = 180^\circ$ (dashed line). For this configuration of opposite winds, the wavelength is found to be remarkably independent of $V/V_{\textrm{\emph{th}}}$ (Fig.~\ref{Wavelength_exp}$(a)$), as opposed to the increase close to the transport threshold observed for $\theta=0^\circ$. Finally, the variation of the characteristic time for pattern formation close to the transition angle is also recovered by the predicted variation of $1/\sigma_{\textrm{\emph{sat}}}$ for divergence angles smaller than $90^\circ$ (figure~\ref{Wavelength_exp}$(b)$). However, this is not the case for $\theta > 90^\circ$.

These results question the ability to observe the linear regime in the experiment, especially when $\theta > 90^\circ$. Such a regime is theoretically associated with bedforms whose wavelength remains constant while their amplitude grows at a rate $\sigma$. Clearly, as soon as a pattern wavelength is detected by autocorrelation of the experimental pictures, it keeps increasing with time, especially for longitudinal dunes (figure~\ref{Image_analysis}$(e,f)$). Enhanced nonlinear effects when $\theta > 90^\circ$ can be understood as follows. In a bidirectional flow regime, transverse dunes are blown from the same side by the two winds. On the contrary, each wind blows from a different side of the longitudinal dunes. As a consequence, the sand flux is all the more affected if the incipient dunes develop an avalanche face. A larger dune aspect ratio is a stronger perturbation for the flow, leading to larger sediment flux variations, and thus an enhanced dune growth rate. These nonlinearities may explain the shortcomings of the linear analysis in reproducing the experimental wavelengths for $\theta > 90^\circ$. These effects also support the shorter characteristic time for dune formation in the longitudinal mode and the prevalence of longitudinal dunes over transverse ones at the transition, when the divergence angle is close to $90^\circ$.

Finally, we can provide estimations of time scales in the experiment, taking as a typical example the case $V/V_{\textrm{\emph{th}}} = 1.4$. Each blow takes $\simeq 4$~s. Because most of the grain motion occurs during the acceleration phase of the plate, the time during which the dune pattern is active is a fraction of the blow duration. The effective wind period $t_1+t_2$ is thus of the order of a few seconds. For $V/V_{\textrm{\emph{th}}} = 1.4$, we have $Q_{\textrm{\emph{exp}}}  = 2~\textrm{mm}^2~\textrm{s}^{-1}$ (figure~\ref{Exp_transportlaw}~$(b)$). Importantly, this is the flux at the crest of barchan dunes. Taking $\gamma = 1.7$ (see section~\ref{ModelCourrech}), we then obtain a reference flux $Q \simeq 0.74~\textrm{mm}^{2}~\textrm{s}^{-1}$. With a saturation length $L_{\textrm{\emph{sat}}} = 1.1~\textrm{mm}$, we have a characteristic pattern growth time  $(1/\sigma_{\textrm{\emph{sat}}})L_{\textrm{\emph{sat}}} ^2/Q \simeq 5$~s, where $\sigma_{\textrm{\emph{sat}}} \simeq 0.35$ is the dimensionless growth rate given by the linear stability analysis corresponding to the experimental conditions ($\mathcal{A}_0 = 4$, $\mathcal{B}_0 = 4$ and $u_*/u_{\textrm{\emph{th}}} = 2.5$). Despite both time scales being comparable in the experiment, the emerging dunes still develop by integrating the two flow directions. While the experimental conditions are at the limit of validity of the linear stability analysis, some of the characteristics of the linear regime remain, such as the variations of the dune orientation or wavelength with respect to the divergence angle.

\section{Concluding remarks}
\label{conclusion}

This work addresses the emergence of a dune pattern from a flat bed submitted to a bidirectional wind regime. It generalises the dimensional approach of \citet{Rubi87} and \citet{Cour14} with a more fundamental linear stability analysis where hydrodynamics over a modulated bed as well as transient sediment transport are described. Our predictions essentially agree with previous results for dune orientation with a transition from transverse to longitudinal dunes, but provide a wider range of possible alignments, especially when the two winds have a divergence angle close to the transition value $90^{\circ}$, depending on flow strength. This analysis also allows us to predict a preferred pattern wavelength, and we show that it either decreases close to the transition for strong winds, due to a geometric effect, or increases at low winds, when the bed slope affects the transport threshold.

In addition, analogous subaqueous experiments where incipient dunes are submitted to alternate forcing in different directions validate part of the model. The geometric effect responsible for the variation of the wavelength close to the transition angle is recovered as well as the increase of the wavelength close to the transport threshold for transverse bedforms. However, longitudinal patterns deviate from the model as they systematically form at smaller wavelengths than expected, and also do not show any variation of their wavelength with the distance to the threshold. Finally, these laboratory experiments show the coexistence of the two pattern orientations predicted by the linear stability analysis close to the transition angle $\theta = 90^{\circ}$. This coexistence continues beyond the linear regime in the experiments but also at much larger scale in deserts on Earth.

To describe the formation of dune patterns under natural multidirectional flow conditions, the analysis can be generalised to any wind rose by adding all the sediment flux contributions. Following \eqref{Sigma_bidi}, the dimensional overall growth rate $\sigma_{\Sigma}$ of a mode $(\alpha,k)$ can be written as
\begin{equation}
\label{Sigma_multi}
        \sigma_{\Sigma}(\alpha,k) = \frac{1}{L_{\textrm{\emph{sat}}} ^2}\frac{\sum_i t_i Q_i \sigma(Q_i,\alpha_i,k)}{\sum_i t_i} \, ,
\end{equation}
where $t_i$ and $Q_i$ are the durations and the fluxes associated with each wind, as derived from field atmospheric measurements. Additional data, such as the typical grain size of the sand bed in order to estimate the velocity threshold $u_{\textrm{\emph{th}}}$ and the saturation length $L_{  sat}$, are also necessary to compute $\sigma_{\Sigma}$. Its maximum then gives the most unstable mode $(\alpha_{\textrm{\emph{sat}}},k_{\textrm{\emph{sat}}})$, predicting the incipient dune orientation and wavenumber for the considered region. As typical sand fluxes in terrestrial deserts are of the order of $Q \simeq 10~\textrm{m}^2~\textrm{yr}^{-1}$, the corresponding characteristic dune growth time $(1/\sigma_{\textrm{\emph{sat}}})L_{\textrm{\emph{sat}}} ^2/Q$ (with $\sigma_{  max} \simeq 0.07$, see figure~\ref{sigma_bidi_maps}) is not much larger than but rather comparable to the seasonal period of wind reorientation (as for our subaqueous experiment). This estimation seems, however, to agree with field observations, as illustrated in figure~\ref{Field_figure}, which emphasises the robustness of the orientation prediction. Despite the difficulties in documenting the linear regime in natural environments, more sites should be systematically tested.

The comparison of the dune orientation predictions with those of \citet{Cour14} suggests that these predictions are still valid in a nonlinear regime as long as dunes develop in areas of high sediment availability. As for the dune wavelength, however, pattern coarsening usually occurs soon after dune emergence and the wavelength grows by means of collisions and coalescence. Driven by the defects in the pattern \citep{Gao15,Day18}, this coarsening is likely to depend on the wind regime. Further studies are needed to investigate this process in order to fully understand the observed complexity of dune fields in nature.

\acknowledgments
We thank O. Devauchelle for discussions and DigitalGlobe for providing the satellite images used in figure~\ref{Field_figure}.

\appendix

\section{Dune growth rate in the model of \citet{Cour14}}
\label{Deriv_cour}
We derive in this appendix the dune growth rate following the dimensional analysis of \citet{Cour14}, here expressed with our notation and conventions. Considering an infinitely long linear dune of height $Z$ and width $W$, the fluxes associated with each of the two winds can be expressed as
\begin{eqnarray}
\boldsymbol{Q}_{1} &=& Q_0\left(1 + \beta\frac{Z}{W_1}\right)\boldsymbol{e}_1 = Q_0\left(1 + \gamma\cos\alpha_1\right)\boldsymbol{e}_1, \\
\boldsymbol{Q}_{2} &=& Q_0\left(1 + \beta\frac{Z}{W_2}\right)\boldsymbol{e}_2 = Q_0\left(1 + \gamma\cos\alpha_2\right)\boldsymbol{e}_2,
\end{eqnarray}
where $W_\textrm{1,2} = W/\cos\alpha_\textrm{1,2}$ are the widths of the dune in the direction of each of the winds. The flux-up ratio $\gamma$ linearly depends on the dune aspect ratio $Z/W$ with a prefactor $\beta$ and accounts for the wind increase at the dune crest. Following the mass conservation equation \eqref{Exner}, the dune growth rates associated with each of the two winds are related to the divergence of those fluxes,
\begin{eqnarray}
\sigma_1 &=& - \frac{1}{Z}\nabla\cdot \boldsymbol{Q}_1 \approx \frac{1}{Z}\frac{Q_1}{W_1}, \\
\sigma_2 &=& - \frac{1}{Z}\nabla\cdot \boldsymbol{Q}_2 \approx \frac{1}{Z}\frac{Q_2}{W_2},
\end{eqnarray}
so that the growth rate $\sigma$ can be expressed as the weighted sum of these contributions:
\begin{eqnarray}
\sigma &=& \frac{1}{N+1}\left(N\sigma_1 + \sigma_2\right) \\
&\approx & \frac{Q_0}{ZW(N+1)}\left(N\cos\alpha_1 + N\gamma\cos^2\alpha_1 + \cos\alpha_2 + \gamma\cos^2\alpha_2 \right).
\end{eqnarray}
%

\bibliographystyle{jfm}

\end{document}